# Project RAINBOW: An all-integrated all-optical ultrafast dual-comb chip

**Author: Tushar Malica [tushar.malica@vub.be, dr.t.malica@gmail.com]**
**Affiliation: Brussel Photonics Team, Dept. of Applied Physics & Photonics, Vrije Universiteit Brussel, Belgium.**

Typically, a laser emits steady light at a single frequency. Mode locking can reshape this light into short bursts of energy, or pulses by forcing constructive interference of the standing waves inside the laser cavity. Contrary to the continuous laser emission, pulsed lasers store multiple frequencies, and all begin in-phase[1,2]. Thus, mode-locked pulses are a fundamental method in ultrafast laser physics to generate a periodic train of phase-coherent optical pulses. In the frequency domain, these form an evenly spaced comb of frequency lines acting like an optical ruler and referred to as an optical frequency 'comb'.

Reducing the pulse duration increases the range of its comprising frequencies[2] leading to the desirable broadening of the associated comb. A broad comb[3,4] can 1) measure and control light like radio waves, and 2) transfer frequency and time signals precisely over a long distance. This principle was awarded the 2005 Nobel Prize in Physics[4] which proposed combs using a femtosecond pulsed laser for applications5 in optical atomic clocks, high-speed data communication[6], and precision laser spectroscopy[7–9]. This was pushed further by the work recognised by the 2018[2] Nobel Prize in Physics to further shorten pulse durations down to femto/attoseconds while simultaneously amplifying the pulse and maintaining its shape. Ultrafast is the shortest timescale known in science and deals with light at the most fundamental level.

Simultaneously, there is an ongoing downsizing and integration of several lasers and optical circuit components to a compact all-integrated photonic chip[5,7,8] that can fit on a fingertip and has a few millimetres dimensions. Photonic integrated chip is a well-established technology incorporating several optical components while saving 10 times power and being 100 times smaller, lighter and cheaper compared to the free-space optics alternatives[10]. These advantages, along with the ability to mass-manufacture make it an attractive candidate to provide state-of-the-art proof-of-concept devices resulting in novel scientific discoveries, patents, and product commercialization.

The first all-on-chip single-pulsed source was demonstrated recently in 2023[11] with a pulse duration of ~4.3 ps at ~10 GHz using nanophotonic-based $LiNbO_3$ hybrid platform. The first demonstration of integrating two separate mode-locked lasers was recently done in 2022 as a multi-chip assembly grown on an InP substrate for LiDAR applications[12]. The current state[5,7,8] of the technology discussed remains predominantly research-centric with research occupying ~93% of the global market in 2022 and projected to a significant ~89% until 2028[13]. This is due to its costly, bulky, partially off-the-chip, and power-hungry nature. Ultrafast comb chip technology occupies a 99% share of the optical frequency comb global market valued at 33mil USD and with a projected rise to 41mil USD by 2028[13].

Project RAINBOW combines ultrafast laser physics, applied physics, and electronic engineering to build an all-integrated standalone photonic chip with an ability to emit two combs (or train of pulses), each at slightly different pulse repetition frequencies using one laser. This proof-of-concept project has neither been demonstrated at any wavelength nor as a single all-optical chip to date. These combs are: 1) phase-locked to each other, and 2) frequency-tuneable with seamless power redistribution between the two combs.

**Scientific objectives**

**(i)** To demonstrate for the first time a functional chip with dual optical frequency comb exhibiting femtosecond pulse durations as shown in fig. 1. The chip will operate around the C-Band telecom wavelengths of ~1550 nm, thus, optimised for communication applications. Ideally, it will have one explicit control with minimal response time for tuneability and maintain minimal chip footprint, power usage, and robustness for mass-manufacturing. Of all the possible ways8 to generate combs such as gain-switched lasers, χ(2) and χ(3) nonlinearities, and electro-optics, passive and hybrid mode-locking techniques [See Tasks1.1 & 2.1] will be used as they inherently support generation of shortest pulses and high optical efficiency. The former has the added advantage of not requiring an additional AC source and external pump, thus, supporting an all-on-one-chip technology.

**(ii)** Explore laser dynamics to enhance chip performance in terms of pulse quality and durations while maintaining a minimal chip footprint. It's done for two different cavity designs as seen in fig.1.

**(iii)** Exploit the scientific observations, in both, the periodically stable pulsed states (mode-locking) and rarely observed novel nonlinear dynamical states such as chimeras. The first set of chip designs show promising results in simulations performed using foundry-recommended Ansys Lumerical licensed software in accordance with rigorous constraints and specifications that are theoretically and experimentally proven[14,15]. Note that these are qualitative results as softwares don't match exact industrial parameters and the design takes additional physical space for metal contacts/ pathways that can't be simulated [See WP1.]

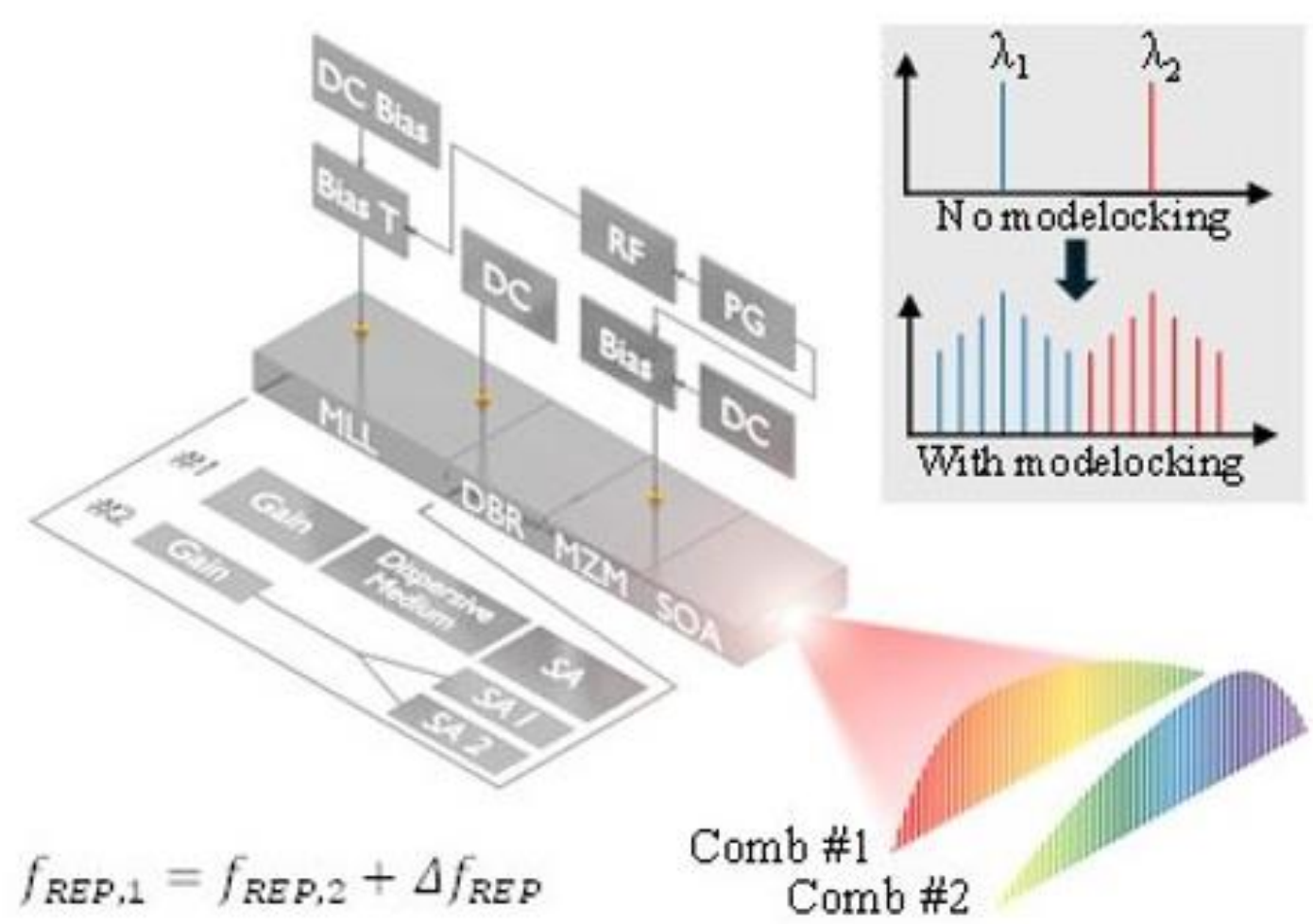

**Fig.1.** Schematic of the ultrafast dual-comb all-integrated photonic chip on InP substrate with two different cavity designs. Key: SA: Saturable absorber; PG: Pattern generator; MZM: Mach-Zehnder modulator; SOA: Semiconductor optical amplifier; DBR; Distributed Bragg reflector.

**Research hypotheses**

**(i)** Here, both combs are coherent. Therefore, all modes of a comb have a characteristic fixed phase relationship, also known as phase-locking. Note that this does not automatically imply that both combs have a phase relationship with each other. Thus, the combs are characteristically exclusive to each other. Locking two spectra to each other such that they are co-dependent to each other, resulting in the associated phases being in synergy, has been a long-standing challenge in the field of ultrafast laser physics. To date, only noise correlation has been observed[16]. I hypothesise that using the same gain medium (and thus the same carrier reservoir) to generate the two combs will facilitate inter-band phase locking. Consequently, one may use this technique to create one "super" comb with an extended frequency comb span comprising the two former combs. This path of investigation will feed into the strategy of tailoring and manipulating comb properties.

**(ii)** Chimeras are observed when there is a non-locally coupled oscillator. A widely accepted definition of chimeras is "a dynamical state where mutually coherent oscillators co-exist with incoherent drifting oscillators"[17,18]. Here, each mode in the comb is an oscillator when passively mode-locked[19]. I can force the device externally or by design to function such that there is a co-existence of mutually coherent modes (phase-locked) and incoherent modes (no fixed relationship between different modes). Note that this is not the same as being partially mode-locked. It will create an intermediate state sustained in time while the system trajectory is on the route to destabilization when the system is neither exhibiting stable behaviour nor completely destabilized. These are intensively studied theoretically, but there is very little experimental evidence of these states to date across all non-linear systems and not just limited to lasers. There is at least a decade long gap between theory and experimental findings of chimeras. Specifically, observing chimeras arising inherently without explicitly adding nonlinearity forcefully, artificially, and externally is rare[19,20]. WP3 and Task3.3 will address this.

**(iii)** At the engineering level, chip demonstration is novel and its functionality will be tested.

**Innovative aspects & breakthroughs**

The innovation lies in converting fundamental science to a first-time technology demonstration. From the perspective of continuous emission multi-wavelength chip devices, recently an all-integrated multi-wavelength continuous laser chip[16,21] using one laser to emit two wavelengths controlled by phase-controlled optical feedback was demonstrated. However, dual-comb with similar specifications as stated here have not yet bee demonstrated to the best of my knowledge. These demosntrated wavelengths will show complete optical power switching between the two modes within a few nanoseconds, grown on an indium phosphide (InP) substrate at ~1550 nm using a generic foundry multi-project wafer approach. Additionally, extending the investigation outside the expected steady-pulsed states towards nonlinear behaviour for observing chimeras will give insights on deterministic mechanism present in many technological and biological self-organised high-dimensional systems such as lasers, neurological workings of the brain, and turbulence in hydrodynamics[22]. Thus, RAINBOW addresses science at two levels: the applied technological level and fundamental physics. Furthermore, the breakthroughs lie in:

(i) demonstrating one stand-alone chip without the need for free-space optics with customised features as stated in the scientific objectives.

(ii) forcing the two wavelengths to mode-lock while managing nonlinear effects that are magnified at such small scale.

(iii) maintaining the pulsed-states overtime over a broad range of applied currents and temperature while trying to reach state-of-the-art femtosecond pulse durations.

(iv) experimental demonstration of "real" chimeras inherent to the laser.

(v) downsizing the technology from free-space optics to an integrated chip leading to a patented design and is a step towards a market-friendly product.

**Research methodology**

RAINBOW comprises three scientific work packages (WPs), each designed to master one independent scientific aspect. The first work package [WP1] aims to simulate and design photonic chips comprising numerous lasers, optical, electrical, and electronic components using a circuit simulator and design software. The second work package [WP2] will demonstrate proof-of-concept and characterize the manufactured photonic chips using an in-house state-of-the-art data acquisition system. The third work package [WP3] will focus on improving the system performance by exploring viable options such as free-space optical engineering and a variety of semiconductor growth platforms for chip integration and enhancing pulse properties. WP3 will also deal with the case when the combs are incoherent, i.e., the phase difference between modes is not fixed, and the lasers are neither fully nor partially mode-locked.

**[WP1]- Photonic Integrated Chip design**

Photonic integration technology comes with a caveat of the existing disconnect between the ecosystems of the fabrication process and the chip design. This leads to the failure of achieving 100% qualitative and quantitative matching between the simulated chip design and the fabricated product to get a “first-time-right” product. Thus, WP1 must run throughout the fellowship. The project demands a few sequential life cycles to achieve the end target; each designed to supersede the performance of older-generation chips. WP1 is a simulation-heavy package split into two tasks: circuit simulation [Task 1.1] and chip layout [Task 1.2]. The outlined strategies below are conventional; however, the chip design and the project’s goal are novel.

**[Task 1.1]** will use Ansys Lumerical INTERCONNECT PIC Simulator. The choice of software is defined by the generic fabrication foundry, Smart Photonics B.V., with whom we have a long-standing relationship. They have a pre-defined library of the basic building blocks following the industrial specification to facilitate the aforementioned caveat.

In parallel, a theoretical multi-mode differential rate-equation model will run on MatLab to simulate the laser’s electric field behavior and increase design accuracy. This strategy will address chip design at both levels- the photonic integration engineering and fundamental laser physics.

The schematic is shown in fig. 1. The intracavity laser light is reshaped into periodic pulses using a saturable absorber (an optical-intensity-dependent element engineered to cyclically attenuate and amplify light). It is placed at the electric field’s anti-node to optimise pulse peak power and maintain pulse symmetry. More importantly, I want only one pulse from each comb inside the laser cavity at any given instant in time, i.e., the fundamental mode-locking regime [See Task2.1] ensuring I am creating a copy of a single pulse and automatically preserving phase coherency and, thus, the comb quality.

Next, I will implement dual-comb generation using two designs, i.e., in series (cavity #1) and in parallel (cavity #2) as shown in fig. 1. Cavity design #1 will create a spatial lag between the two combs using a chromatic dispersive waveguide. While cavity design #2 uses two separate saturable absorbers that will work independently for each comb.

I specify spatial separation between the central peak frequency (or wavelength) of the combs by $\Delta f_{REP}$=30 GHz (or 10nm) bearing in mind the fabrication precision limit. The pulses will pass the DBR layer optimally matching the peak emission wavelengths, followed by modulation through MZM and PG offering flexibility in pulse repetition frequencies. Isolators will be used, as required, to prevent back reflections. Finally, SOA amplifies pulse energy. For cavity design #2, I will reduce the chip footprint by using a conventional idea of folding the cavity at an angle of 5°-7°.

**[Task 1.2]** will use results from Task1.1 to prepare a chip blueprint known as the mask layout using Luceda IPKISS. The time utilized for Task1.2 also bears in mind that there are constraints such as accounting for the optical losses and the physical space used for metal connectors between the chip components to ensure the desired simulated design translates optimally on the fabricated chip. At the end of each design lifecycle, the blueprint will be outsourced to Smart Photonics B.V., an Eindhoven-based generic foundry in the Netherlands. The selection of foundry is based on the chosen chip growth platform, i.e., InP because it 1) inherently works best at ~1550nm, and 2) one of the commercially available option allows for a 100% on-chip device with universal integration of waveguides, lasers, active and passive photonic components[7,8,23].

**[WP2]-Experiment**

WP2 will kick-off once the first-generation chips are delivered estimated within the first 6 months of the fellowship. WP2 is an experiment-heavy package and exploits fabricated chips through two tasks, namely, [Task 2.1] for summarising the parameters affecting system performance via systematic characterization and [Task 2.2] will implement phase stabilization of individual combs and phase-locking between the two combs simultaneously.

**[Task 2.1]** will 1) inform me of explicit chip parameters affecting system performance and sensitive to phase change to advance to Task2.2, and 2) help update the blueprint of new generation chips designed concurrently in WP1. All chips will be characterized through an in-house nanoscale-controlled probe station. Temporal and spectral measurements will be performed using high-resolution optical analyzer (APEX AP2083, 40 fs resolution), autocorrelator (Femtochrome FR-103XL ~2fs resolution), RF spectrum analyzer (Keysight N9021B, 42 GHz) and oscilloscope (Teledyne LabMaster 10 Zi-A oscilloscope, 20 GHz bw, 80 GS/s); the two last ones being combined with a high-bandwidth photoreceiver (ThorLabs RXM40AF, 42 GHz). The chip will be functional in two operational modes for both laser cavity designs as shown in fig.1: 1) Passively mode-locked with current applied to the gain and SA either switched off or having a small bias voltage, and 2) Hybrid mode-locked when the gain is driven by a direct current (DC), and the saturable absorber is driven by a bias tee with a reverse DC voltage plus RF modulation detuned away from the pulse repetition frequency.

Rationale is as follows: Chips intrinsically support mode-locking as they have low propagation losses with negligible phase noise and laser chirp over long distances and maintain the narrow linewidth over extended cavity lasers. It is well-established that passive mode-locking with a saturable absorber operating in the fundamental regime 1) generates the shortest optical pulse, 2) self-starts the mode-locking mechanism, 3) allows only one pulse inside the laser cavity with each output pulse being a copy of the same pulse causing minimal pulse-to-pulse variation and phase, and 4) passive elements allow for a minimal fabrication error and robustness for mass manufacturing. Meanwhile, active mode-locking is intentionally avoided here as the intracavity elements can't be individually accessed in isolation at such dimensions to trigger precise active modulation of intracavity loss. However, hybrid mode-locking gives me the freedom to exploit the advantages of modulating specific experimental parameters and capitalize on the modulation properties without trading off device performance.

A basic characterization of all chips will be performed to give a general performance overview to choose 3-4 best-performing chips which will be packaged for long-term use. This is done because not all fabricated chips, although grown on the same III-V semiconductor wafer and with identical designs, will perform the same due to 1) the limiting resolution of industrial fabrication processes and different locations of chips on a semiconductor wafer, and 2) to keep packaging-related costs low. The chip packaging is outsourced to PhotonFirst B.V. (Netherlands) and takes 3-4 weeks. The packaged chips will undergo detailed systematic characterisation and be used for the next steps.

**[Task 2.2]** Ideally, an ultrafast laser has an ultra-broad frequency comb spanning a few Terahertz. In practice, dispersion occurs due to the differences in the associated group ($\upsilon_g$) and phase ($\upsilon_p$) velocities leading to an offset between the peak of the envelope and the closest peak of the carrier wave known as the carrier-envelope offset3 with an associated phase of $\theta=[1/\upsilon_g -1/\upsilon_p]L\omega(\text{mod}2\pi)$, where L is the laser cavity length and ω is the carrier frequency. The carrier offset frequency of each comb ($f_o$) is directly proportional to this θ. This impacts the entire comb since each frequency comb line is born from $f_o$.

Thus, WP2.2 aims to optimize device performance by solving/alleviating this fundamental issue by 1) locking individual combs to an external reference laser source [Keysight N7776C tuneable laser source], and 2) active stabilization. Both involve a traditional off-the-chip setup extended in free-space to gain knowledge of the physics involved in the interaction of the frequency combs. To date, researchers have only observed noise correlation in a muti-wavelength device and never achieved intrinsic phase-locking[16]. Not to forget that management of optical nonlinearity at such dimensions and ultrafast scale amplifies the magnitude of the issue I am addressing. Hence, every insight gained in Task 2.2 will be a novel and fundamental contribution to the field of laser physics. Pushing the envelope further, I will scale these results down to integrate into the later-generation chip designs in WP1. Taking into consideration that real-time phase measurement at an ultrafast scale is complex, time-consuming to build, and expensive; I will use the traditional technique of measuring the beat note between the two combs using our photodetector. This figure of merit is an indicative of the quality of mode locking and will achieve the same quality results without the need for an expensive time-consuming real-time high-resolution single-shot aforementioned phase measurement.

**[WP3]-Femtosecond Pulse generation**

The chip thus far uses an industrial standard InP-based edge-emitting laser with 3 quantum wells. A ~5 ps pulse duration is expected, mainly limited by the saturable absorber recovery time of 4 ps[24]. Realistically, the first-generation chip will have a pulse duration of 15-30 ps considering all electrical and optical losses. To break this barrier and enter the femtosecond regime, [Task 3.1] will shorten the pulses off-the-chip, and [Task 3.2] explore options to redesign the laser and integrate results from Task 3.1. [Task 3.3] will dive into specific operating conditions and laser parameters that will create chimeras inherent to the system.

**[Task 3.1]** I will use the most popular technique to compress picosecond duration pulses in free-space optics, i.e., using two diffraction gratings[25,26]. Each grating will be engineered by choosing the angled position, length, and the material (i.e. effective refractive index) to shape the leading and trailing edges of the pulse individually for optimal dispersion compensation. Additionally, mirrors with slightly different peak reflectivities can be used to support the different spectral ranges of the two frequency combs. Few constraints must be considered: 1) the ratio of average output power to the pulse repetition frequency gives the pulse peak power, and 2) an ideal symmetric hyperbolic secant pulse has a peak power of 0.88 times the ratio of the pulse energy to the pulse duration. Not to forget, nonlinear processes such as higher harmonic generation[2] become significant. Thus, ensuring that each pulse can hold the intracavity optical power in the femtosecond regime optimally without any side effects such as a change of central wavelength (chirping[2]) and pulse stretching/splitting is the focus of this task. To intrinsically stabilise the system, I will 1) fix the pulse repetition frequency at 10 GHz as III-V semiconductor gain material has a short upper state carrier lifetime of a few nanoseconds[27]. This will also fix the frequency spacing of the adjacent frequencies in the comb. 2) The two frequency combs' wavelengths will be spaced 10 nm apart to avoid unwanted nonlinear interactions. 3) The difference between two pulse repetition frequencies will arise from the difference in the optical path length[3]. This will

also alleviate the carrier-envelope offset phase discussed in Task2.2. Once the setup is optmised in free-space, I will use the results obtained to scale it down and incorporate it in the revised chip design.

**[Task 3.2]** Recent technological advances allow us to exploit the advantages of different semiconductors and combine platform technologies, i.e., hybrid and heterogeneous integration. Besides InP platform, en masse production of photonic chips is popularly done on silicon nitride (SiN) and silicon (Si) platforms. More recently, nonlinear-medium-on-insulator[7] platforms have shown up using Lithium niobate ($LiNbO_3$) and Aluminium gallium arsenide (AlGaAs). My choice of InP platform up to Task3.1 was dictated by it being the sole option offering full integration of all photonic components [See Task1.2]. Si-based options don't support the idea of "all-on-chip" as its physical properties prohibit optical gain for laser or amplifier, thereby, needing external optics and increasing coupling losses and power usage. Hence, this task is purely exploratory and intended to optimize long-term impact, exploitation, and incorporate results from Task3.1.

**[Task 3.3]** Upto this point, the project deals with coherent combs, i.e., all modes are phase locked with fixed relative phase constants. In this task, I will answer the question: What happens if there is a coexistence of fully/partially mode-locked comb and an incoherent comb? It has been previously concluded that experimental observation of system-inherent chimeras has two prerequisities in mode-locked lasers, i.e., the choice of material and the device length[19]. Presence of quantum dots (QDs) encourages this state due to the possibility of pulses with broad trailing edge plateau (TEP) leading to decelerating carrier exchange between the career reservoirs and ground states of the QDs[28]. This acts as a filter and homogenizes the carrier and photon distribution along the gain material. The trade-off is that the pulses will be assymetrical with elongation of the trailing edge. Note that there is no replacement for QDs in quantum-well devices as the carrier-photon saturation is not the same[19]. Another way is by inherent laser design, i.e., shortening the length of the gain medium such that carrier-photon saturation occurs and mimics TEP. Here, I am limited by the industrial standards of chip manufacturing. Both of these lines of investigation will be performed in this task and are heavily dependent on findings of tasks 3.2 & 2.1. The main difference of this task as opposed to previous tasks will be that the end goal is intentionally drive the system away from stabilization and modifying the laser design via the two aforementioned parameters. This task looks for the optimal set of operating conditions where the system exhibits co-existing coherent and incoherent spectra using the same data acquisition system as stated in task 2.1.